\documentclass{article}
\usepackage{spconf,amsmath,graphicx}
\usepackage{amsmath,graphicx}
\usepackage{xspace}
\usepackage{siunitx}
\usepackage{bm}
\usepackage{hyperref}
\usepackage[justification=centering]{subcaption}
\usepackage{fancyhdr}

\usepackage{enumitem}
\setlist{nosep, leftmargin=14pt}

\usepackage{mwe} 
\usepackage{amsfonts}

\graphicspath{{images/}}

\newcommand{\xset}[1]{\ensuremath{\mathcal{X}_{#1}}\xspace}

\newcommand{\xtest}{\xset{te}}
\newcommand{\xtrans}{\xset{un}}

\newcommand{\scratch}{\texttt{scratch}\xspace}
\newcommand{\distance}{\texttt{distance}\xspace}
\newcommand{\con}{\texttt{contrastive}\xspace}

\usepackage{xcolor}

\newcommand{\fref}[1]{Fig.~\ref{#1}}
\newcommand{\tref}[1]{Table~\ref{#1}}

\newcommand{\mm}{\si{\milli\metre}}
\newcommand{\micron}{\si{\micro\metre}}

\fancypagestyle{specialfooter}{%
  \fancyhf{}
  
  \fancyfoot[C]{\textbf{$\copyright$ 2021 IEEE. Personal use of this material is permitted. Permission from IEEE must be obtained for all other uses, in any current or future media, including reprinting/republishing this material for advertising or promotional purposes, creating new collective works, for resale or redistribution to servers or lists, or reuse of any copyrighted component of this work in other works.}}
}
\title{Contrastive Representation Learning for Whole Brain Cytoarchitectonic Mapping in histological Human Brain Sections}

\name{Christian Schiffer$^{1,2}$ \qquad Katrin Amunts$^{1,3}$ \qquad Stefan Harmeling$^{4}$ \qquad Timo Dickscheid$^{1,2}$}
\address{
	$^{1}$ Institute of Neuroscience and Medicine (INM-1), Research Centre Jülich, Germany\\
	$^{2}$ Helmholtz AI, Research Centre Jülich, Germany\\
	$^{3}$ C\'{e}cile \& Oscar Vogt Institute for Brain Research, University Hospital Düsseldorf, Germany\\
	$^{4}$ Institute of Computer Science, Heinrich-Heine-University Düsseldorf, Germany
}

\begin{document}
%
\thispagestyle{specialfooter}
\maketitle
\begin{abstract}
	Cytoarchitectonic maps provide microstructural reference parcellations of the brain, describing its organization in terms of the spatial arrangement of neuronal cell bodies as measured from histological tissue sections.
	Recent work provided the first automatic segmentations of cytoarchitectonic areas in the visual system using Convolutional Neural Networks.
	We aim to extend this approach to become applicable to a wider range of brain areas, envisioning a solution for mapping the complete human brain.
	Inspired by recent success in image classification, we propose a contrastive learning objective for encoding microscopic image patches into robust microstructural features, which are efficient for cytoarchitectonic area classification.
	We show that a model pre-trained using this learning task outperforms a model trained from scratch, as well as a model pre-trained on a recently proposed auxiliary task.
	We perform cluster analysis in the feature space to show that the learned representations form anatomically meaningful groups.
\end{abstract}

%
\vspace*{-.5\baselineskip}
\section{Introduction}
\label{sec:intro}
\vspace*{-.5\baselineskip}

The cytoarchitecture of the human cerebral cortex represents an important microstructural reference for structural analysis and localizing signals of brain connectivity and function~\cite{Amunts2015, Amunts2020}.
Cytoarchitectonic areas are characterized by the distribution, arrangement and layering of neuronal cells. They can be identified in cell-body stained histological brain sections, imaged at a resolution of about 1 micrometer using light microscopy.
The most established method for identification of cytoarchitectonic borders (\textit{cytoarchitectonic mapping}) uses statistical analysis of cell segmentations along cortical profiles \cite{Schleicher1999}. It is precise and reliable, but not designed to handle large amounts of data from high throughput imaging in a highly automated fashion. An automatic method that can be applied at high throughput with limited supervision would provide a strong benefit for advancing our understanding of brain structure and function~\cite{Amunts2020}, by allowing to compute maps in more tissue sections and brain samples than possible today, and thus providing a more comprehensive sampling of the structural variability in the brain.

\cite{Spitzer2017, Spitzer2018} formulated cytoarchitectonic mapping as an image segmentation task, and 
presented the first results for different areas of the visual system.
They identified limited availability of training examples as a key problem, and proposed a self-supervised \textit{distance task} \cite{Spitzer2018} on 3D location and distance prediction in the BigBrain model~\cite{Amunts2013} to learn robust cytoarchitectonic features.
The features improved classification performance significantly compared to models that were trained from scratch.

Our aim is to further extend the approach in \cite{Spitzer2018} to achieve classifications for a wider range of cytoarchitectonic areas, going beyond the visual system and across different brain samples.
The considerable variability of cytoarchitectonic areas especially in the human brain makes this a very challenging problem.
Following the strategy of \cite{Spitzer2018}, we aim to pre-train the network to produce robust and highly distinctive representations of cortical image patches.
In particular, we propose a pre-training task based on contrastive learning~\cite{Hadsell2006, Chen2020, Khosla2020},  which allows to exploit most of the manual brain area annotations available in our lab~\cite{Amunts2020} for learning.
We compare the performance of the proposed method to models trained with classical cross-entropy loss from scratch, and from weights obtained in~\cite{Spitzer2018}.

\section{Methods}%
\vspace*{-.5\baselineskip}
\label{sec:methods}
\vspace*{-.5\baselineskip}


\textbf{Contrastive training.}
Contrastive learning~\cite{Hadsell2006} aims to derive a functional mapping from high-dimensional data points (e.g.~images) to lower dimensional feature representations, so that similar data points map to similar features, and dissimilar data points to dissimilar features.
\cite{Chen2020} showed that representations produced by contrastive learning can improve classification performance when fine-tuned on labeled images.
Given an input image, they create two different \textit{views} by applying random transformations, and then train a neural network to produce similar hidden representations for the same source image.
\cite{Khosla2020} further generalize the notion of similarity to all image patches sampled from the same class.
Using their proposed \textit{supervised contrastive loss}, they present improved classification performance compared to training with a common cross entropy loss.

We build on the method of \cite{Khosla2020} to learn robust representations for cytoarchitectonic area classification in the human brain.
Initial experiments showed that contrastive learning based on data augmentation as in \cite{Chen2020, Khosla2020} is not well suited for this task:
While learned representations showed invariance wrt.~transformations applied during training (e.g.~rotation or intensity transformation), they failed to capture the subtle cellular details that actually distinguish brain areas.
In fact the learned features focused on capturing anatomical landmarks such as blood vessels, which are robust under common image transformations but not necessarily descriptive of cytoarchitecture.
To overcome this undesired effect, we chose to define similarity exclusively by two patches having the same class label.

We train an encoder neural network $f\left(\bm{x}\right) = \bm{h}$ to map an image patch $\bm{x}$ to a lower dimensional representation $\bm{h} \in \mathbb{R}^{D_e}$, which is then fed into a projection network $g\left(\bm{h}\right) = \tilde{\bm{z}} \in \mathbb{R}^{D_p}$ to produce a normalized input vector $\bm{z}_i = \tilde{\bm{z}}_i / \left\lVert \tilde{\bm{z}}_i \right\rVert_2$ for the contrastive loss.
Considering a minibatch of $N$ images $\bm{x}_i$ with corresponding labels $y_i$ ($i=1,\ldots,N$), the contrastive loss $\mathcal{L}$ is computed as
\begin{equation}
	\mathcal{L} = \frac{1}{N} \sum_{i=1}^N \mathcal{L}_i
	\label{eq:contrastive_loss}
\end{equation}
\begin{equation}
	\mathcal{L}_i = -\frac{1}{N_{y_i}} \sum_{j=1}^N \mathbb{I}_{i \neq j}  \mathbb{I}_{y_i = y_j} \log \frac{e^{ \langle \bm{z}_i, \bm{z}_j \rangle / \tau} }{\sum_{k=1}^N \mathbb{I}_{i \neq k} e^{ \langle \bm{z}_i, \bm{z}_k \rangle / \tau} }
\end{equation}
%
where $\mathbb{I}$ is the indicator function, $\tau \in \mathbb{R}$ is a temperature scaling parameter, and $N_{y_i}$ is the number of image patches in the batch with the same label as $\bm{x}_i$.
The loss formulation is adapted from \cite{Khosla2020}.
After contrastive training, image classification is performed by discarding the projection network $g$ and training a linear classifier on the feature vectors $\bm{h}$ extracted by $f$ while keeping parameters of $f$ fixed.

\textbf{Model architecture.}
While \cite{Spitzer2018} formulated cytoarchitectonic mapping as a segmentation problem, we study the related patchwise classification problem:
Given an image patch extracted from the center of the cortical ribbon, a neural network is trained to predict the corresponding cytoarchitectonic area.
We design $f$ identical to the encoder of the modified U-Net~\cite{Ronneberger2015} proposed in \cite{Spitzer2017}, which has been applied successfully for classification of visual areas.
The architecture is composed of six blocks of layers, each consisting of two convolutional layers.
All but the last block are followed by $2 \times 2$ max-pooling layers.
The number of filters is identical for all layers within a block and is set to $16, 32, 64, 64, 128$ and $128$, respectively ($D_e = 128$).
The first convolution in the first block uses a filter size of 5 with stride 4, all remaining convolutions use a filter size of 3 with stride 1.
Each convolutional layer is followed by batch normalization~\cite{Ioffe2015} and ReLU activation.
The projection network $g$ is implemented as a multi-layer perceptron with one hidden layer of dimension 128 and output dimension $D_p = 128$. It also uses batch normalization and ReLU activation.

\vspace*{-.5\baselineskip}
\section{Results}%
\label{sec:results}
\vspace*{-.5\baselineskip}

\begin{figure}[t]
	\includegraphics[width=\columnwidth]{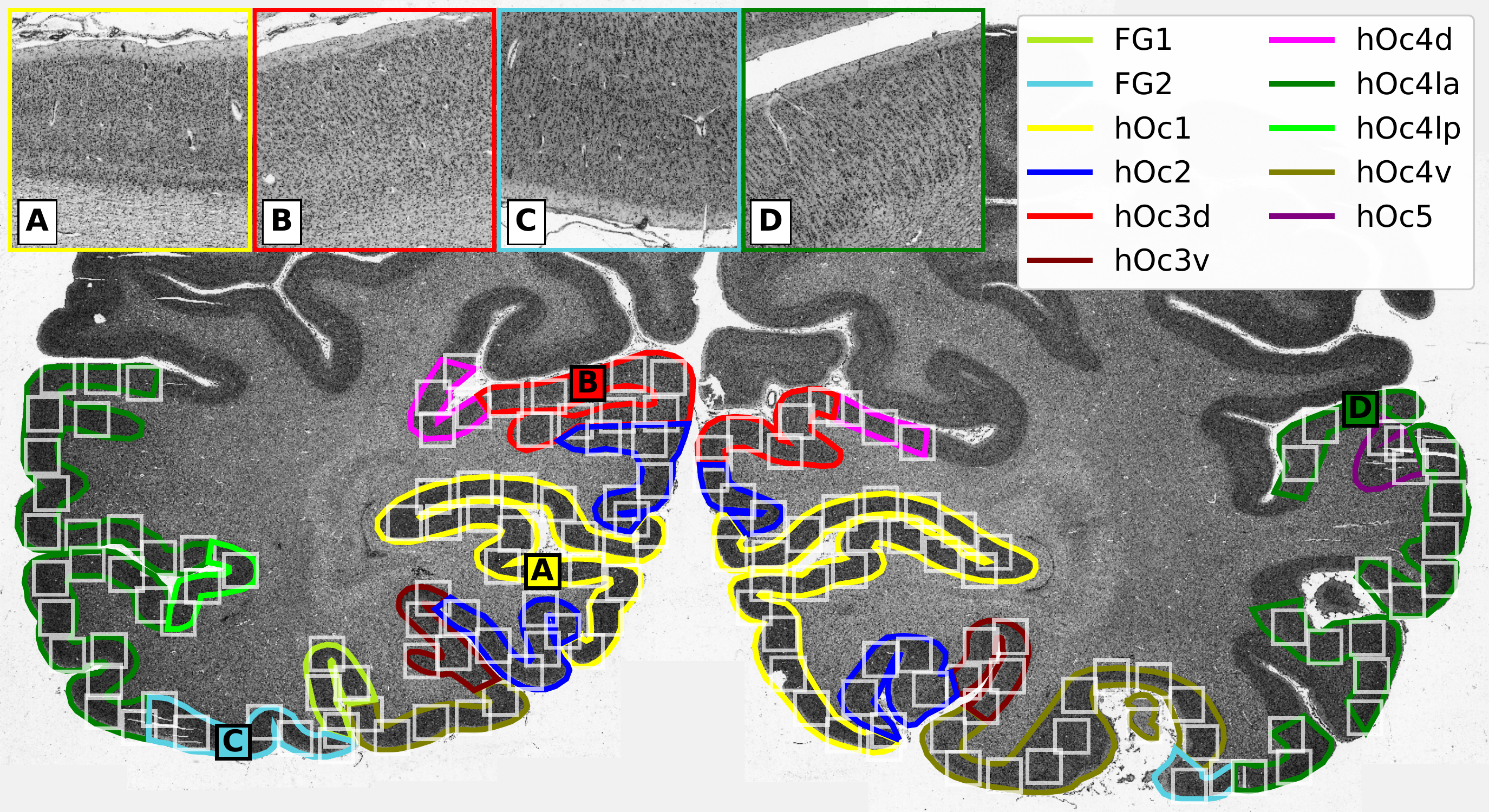}
	\caption{
		High resolution scan of a histological human brain section.
		Training patches (squares) are extracted along a centerline through the cortical ribbon, as derived from the reference delineations of cytoarchitectonic areas (colored contours).
		High resolution example image patches (A-D) show detailed cytoarchitectonic features for patches marked as solid squares.
		Image patches were gamma corrected ($\gamma=0.5$) to improve visualization.
	}
	\label{fig:patches}
\end{figure}

\textbf{Dataset.}
The dataset contains 1860 histological human brain sections that were stained for cell bodies and sampled from 7 postmortem human brains. We use annotations of 113 cortical cytoarchitectonic areas from different parts of the brain (e.g. visual, auditory, motor).
For training, 1200 patches ($1129 \times 1129$ pixels at $2 \micron$ per pixel) per cytoarchitectonic area were extracted from $80\%$ of the sections, oversampling smaller areas for class balancing (\fref{fig:patches}).
After training, performance was evaluated on a test set \xtest{} with patches from tissue sections that were excluded from training.
Furthermore, transferability of trained models to a previously unseen brain was evaluated on image patches from 325 sections of another brain, referred to as \xtrans{}.

\textbf{Data augmentation.}
Data augmentation was applied to capture typical variations in the data ($U[a, b]$: uniform distribution over $[a, b]$):
Patches were rotated by $\theta \in U\left[-\pi, +\pi\right]$, patch center positions were translated in random direction by $d \sim U\left[0\mm, 0.2\mm\right]$ and patches were mirrored vertically with a probability of $50\%$.
Pixel intensities  $x \in \left[0, 1\right]$ were randomly augmented using \emph{unbiased gamma augmentation}~\cite{Pohlen2017} $\alpha x^\gamma + \beta$ with parameters $\alpha \in U[0.9, 1.0]$, $\beta \in U[-0.1, +0.1]$, $\gamma = \frac{\log \left(0.5 + 2^{-0.5 Z}\right)}{\log \left(0.5 - 2^{-0.5 Z}\right)}$, $Z \sim U[-0.05, +0.05]$.
Furthermore, images were blurred with an isotropic Gaussian $G_\sigma (x)$ of kernel size $\sigma \sim U[0.125, 1.0]$, or sharpened according to $x + \delta (G_{\sigma_u} (x) - x)$ with ${\sigma}_u \sim U[0.125, 1.0]$, $\delta \sim U[0.5, 1.5]$ with probability $25\%$, respectively.

\textbf{Training.}
Contrastive feature learning was performed for $150$ epochs using the LARS optimizer~\cite{You2017} with constant learning rate $0.01 * N / 128$, batch size of $N = 4096$ and $\tau=0.07$.
We choose a large batch size to allow representation of multiple class instances in one batch, providing many positive and negative samples per class as in \cite{Chen2020}.
The training process was distributed across 32 GPUs (NVidia K80, 12GB) on 8 compute nodes of the JURECA HPC cluster~\cite{Krause2018}.
Batch normalization statistics were averaged across GPUs.
Training was implemented using \textit{PyTorch} \cite{Paszke2019}.
This model is denoted as \con{}.
After training the projection network $g$ was discarded, parameters of $f$ fixed and a linear classifier was trained using cross-entropy loss for 30 additional epochs to map learned representation vectors to cytoarchitectonic areas using the same training parameters as for the contrastive training.
To assess the benefit of contrastive representation learning, we trained the same network architecture using cross-entropy.
We stacked a linear classifier on top of $f$ and trained the model end-to-end for 180 epochs, so the total number of iterations corresponds to the contrastive learning setup.
We trained one model from scratch by randomly initializing the network weights (denoted \scratch{}), and initialized another model from weights obtained with the self-supervised distance task from~\cite{Spitzer2018} (denoted \distance{}).

\textbf{Evaluation.}
We report F1 scores, top-1 and top-3 accuracy obtained on \xtest{} and \xtrans{} by \scratch{}, \distance{} and \con{} (\tref{table:metrics}).
F1 scores are weighted based on label frequency to account for the unbalanced number of patches per area.
We then cluster representation vectors of patches in \xtest{} into $10$ clusters using Ward hierarchical clustering~\cite{WardJr1963},
and analyze in how far structurally similar brain areas group into the same cluster (\tref{table:cluster}).

\con{} outperforms \scratch{} and \distance{} in \xtest and \xtrans in all considered metrics, demonstrating the advantage of contrastive pre-training compared to end-to-end training with cross-entropy.
\distance{} does not obtain better performance compared to \scratch.
Performance on \xtrans{} is lower compared to \xtest{}.

\begin{table}[t]
	\begin{tabular}{llrrr}
\hline
 model                & dataset   &   F1 score &   top-1 &   top-3 \\
\hline
 \texttt{scratch}     & \xtest{}  &      43.31 &   43.16 &   70.6  \\
 \texttt{distance}    & \xtest{}  &      27.27 &   27.64 &   51.51 \\
 \texttt{contrastive} & \xtest{}  &      \textbf{50.49} &   \textbf{50.08} &   \textbf{78.57} \\
\hline
 \texttt{scratch}     & \xtrans{} &      22.85 &   22.63 &   44.23 \\
 \texttt{distance}    & \xtrans{} &      10.77 &    9.79 &   19.63 \\
 \texttt{contrastive} & \xtrans{} &      \textbf{25.03} &   \textbf{23.94} &   \textbf{46.57} \\
\hline
\end{tabular}

	\caption{F1 score, top-1 and top-3 accuracy obtained on \xtest and \xtrans by investigated models.
		\con{} outperforms \scratch{} and \distance{}~\cite{Spitzer2018} on both datasets.
	}
	\label{table:metrics}
\end{table}

\tref{table:cluster} shows the high-dimensional clusters visualized in 2D after dimensionality reduction with \textit{t-SNE} \cite{Hinton2003}.
We see that many structurally and functionally related areas are assigned to the same cluster:
Cluster 8 consists almost exclusively of features from the primary visual cortex (hOc1).
Features in cluster 2 and 10 come mostly from higher visual areas, those in cluster 9 mostly from areas related to somatosensory functions.
Some of the other clusters are dominated by patches from specific brains:
Features in clusters 2 and 10 come from similar areas, but cluster 2 is dominated by two specific brains.
Most features in cluster 10 however originate from three other brains.

\begin{figure}[t]
	\includegraphics[width=\columnwidth]{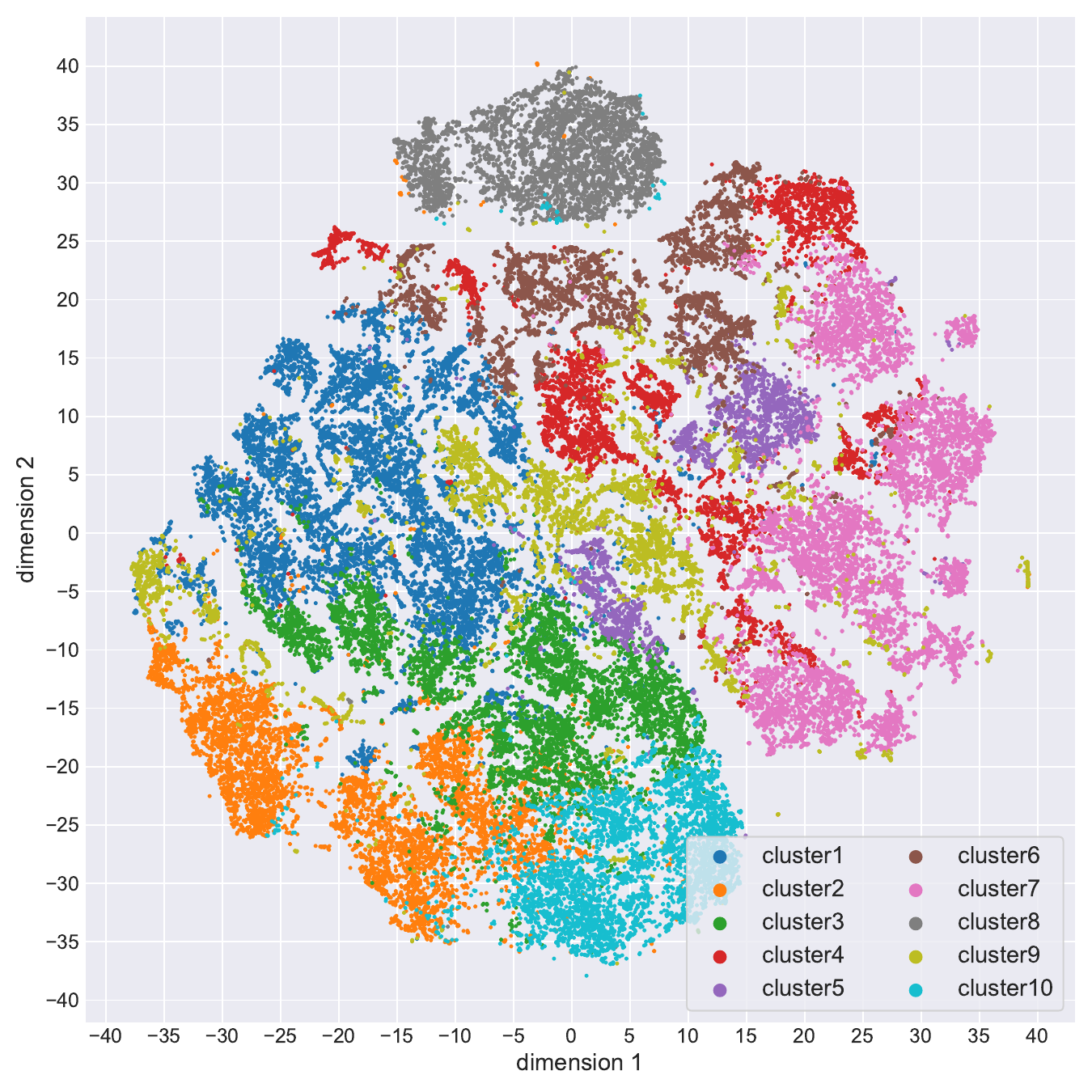}
	\caption{
		Visualization of hierarchically clustered feature representations of \xtest{} in a two-dimensional t-SNE embedding.
		Frequent cytoarchitectonic areas per cluster are given in \tref{table:cluster}.
	}
	\label{fig:clusters}
\end{figure}

\begin{table}[t]
	\begin{tabular}{llll}
\hline
 \#   &                 &                  &                 \\
\hline
 1   & STS2 (9.5\%)     & TE3 (7.7\%)       & STS1 (6.3\%)     \\
     & PF (4.5\%)       & area 1 (3.7\%)    & area 7A (3.6\%)  \\
\hline

 2   & hOc2 (19.2\%)    & hOc4lp (13.7\%)   & hOc3d (12.4\%)   \\
     & hOc4la (12.0\%)  & hOc3v (9.8\%)     & hOc4v (8.7\%)    \\
\hline

 3   & hOc4la (9.3\%)   & FG4 (7.9\%)       & PGp (7.7\%)      \\
     & FG2 (7.0\%)      & hIP5 (6.2\%)      & FG3 (5.6\%)      \\
\hline

 4   & area 45 (17.0\%) & Fo3 (6.9\%)       & Fo6 (6.7\%)      \\
     & area 8c (6.1\%)  & Fo1 (5.3\%)       & area 8d (5.3\%)  \\
\hline

 5   & p32 (7.5\%)      & area 8c (5.0\%)   & area 8d (4.9\%)  \\
     & area 8a (4.0\%)  & area 44 (3.6\%)   & area 8b (3.5\%)  \\
\hline

 6   & p32 (22.4\%)     & p24b (10.6\%)     & Id6 (7.7\%)      \\
     & s32 (6.2\%)      & Fo2 (5.6\%)       & pd24cv (5.5\%)   \\
\hline

 7   & area 4a (10.8\%) & area 6v1 (10.3\%) & area 8a (9.4\%)  \\
     & area 6d2 (9.0\%) & area 6ma (8.7\%)  & area 6v2 (8.3\%) \\
\hline

 8   & hOc1 (98.2\%)    & hOc2 (1.2\%)      & area 5L (0.2\%)  \\
     & area 3b (0.1\%)  & hOc3d (0.1\%)     & hOc4d (0.0\%)    \\
\hline

 9   & area 3b (9.5\%)  & area 1 (6.1\%)    & area 3a (6.1\%)  \\
     & area 5L (5.2\%)  & area 5M (4.5\%)   & STS2 (4.4\%)     \\
\hline

 10  & hOc2 (42.5\%)    & hOc3v (17.7\%)    & hOc4v (10.4\%)   \\
     & hOc3d (10.1\%)   & hOc4lp (4.4\%)    & hOc4d (4.3\%)    \\
\hline
\end{tabular}

	\caption{Percentage of features inside each cluster which belong to image patches from specific cytoarchitectonic areas.
	Only most common areas are listed, therefore the fractions do not sum to $100\%$.
Brain area names follow the nomenclature of the Julich-Brain cytoarchitectonic maps~\cite{Amunts2020}.}
	\label{table:cluster}
\end{table}

\vspace*{-.5\baselineskip}
\section{Discussion \& Conclusion}%
\label{sub:conclusion}
\vspace*{-.5\baselineskip}

We proposed a new strategy based on contrastive learning for encoding cortical image patches from microscopic scans of human brain sections with feature representations that are effective for classifying cytoarchitectonic areas.
In our experiments, models pre-trained using this contrastive loss outperformed models that were trained end-to-end with cross-entropy loss.
The overall precision level as measured by the F1 score is rather low, reflecting the general difficulty in performing very precise classification of cytoarchitectonic areas.
%
Cluster analysis in feature space confirmed that the learned representations capture relevant cytoarchitectonic properties, and often encode areas from related functional or structural regions into the same feature clusters.

To some degree, the learned feature representations seemed to encode specific aspects of particular brains, although the contrastive loss promotes similar patches only to originate from the same brain area, \textit{independent of the brain sample}. This effect is not desired for brain mapping, and future work will investigate ways of overcoming it.
%
Unexpectedly, we could not confirm by experiment that the \distance{} architecture provides better performance than \scratch. This suggests that pre-training on data from a single brain might have little or no benefit for predicting areas from different brains.

While our experiments address patchwise classification as a downstream task, the findings should directly apply to image segmentation as in \cite{Spitzer2018}. 

In future work, we plan to investigate different hyperparameter settings and neural network architectures, as well as optimized methods to further promote learning of subject agnostic features.
Finally, we plan to combine the proposed feature representations with neuroanatomical priors for brain mapping.

\small{
\vspace*{\baselineskip}
\noindent{\textbf{Compliance with ethical standards}
The studies carried out require no ethical approvals.
The human post mortem brains for these studies were obtained by the body donor programs of the Anatomical Institute of University of Düsseldorf, Rostock and Aachen or from collaboration between Anatomical Institute and departments of pathology in accordance with legal requirements.
All body donors have signed a declaration of agreement.

\noindent{\textbf{Acknowledgments}
This project received funding from the European Union’s Horizon 2020 Research and Innovation Programme, grant agreements 785907 (HBP SGA2) and 945539 (HBP SGA3), and from the Helmholtz Association’s Initiative and Networking Fund through the Helmholtz International BigBrain Analytics and Learning Laboratory (HIBALL) under the Helmholtz International Lab grant agreement InterLabs-0015.
Computing time was granted through JARA on the supercomputer JURECA at Jülich Supercomputing Centre (JSC) as part of the project CJINM16.
The authors declare no competing interests.
}
\vfill

\bibliographystyle{IEEEbib}
\bibliography{strings,refs}

\begin{thebibliography}{10}

\bibitem{Amunts2015}
K.~Amunts and K.~Zilles,
\newblock ``Architectonic {{Mapping}} of the {{Human Brain}} beyond
  {{Brodmann}},''
\newblock {\em Neuron}, vol. 88, no. 6, pp. 1086--1107, 2015.

\bibitem{Amunts2020}
K.~Amunts et~al.,
\newblock ``Julich-{{Brain}}: {{A 3D}} probabilistic atlas of the human brain's
  cytoarchitecture,''
\newblock {\em Science}, vol. 369, no. 7, pp. 988--992, 2020.

\bibitem{Schleicher1999}
A.~Schleicher et~al.,
\newblock ``Observer-{{Independent Method}} for {{Microstructural
  Parcellation}} of {{Cerebral Cortex}}: {{A Quantitative Approach}} to
  {{Cytoarchitectonics}},''
\newblock {\em NeuroImage}, vol. 9, no. 1, pp. 165--177, 1999.

\bibitem{Spitzer2017}
H.~Spitzer et~al.,
\newblock ``Parcellation of visual cortex on high-resolution histological brain
  sections using convolutional neural networks,''
\newblock in {\em {{ISBI}}}, 2017, pp. 920--923.

\bibitem{Spitzer2018}
H.~Spitzer et~al.,
\newblock ``Improving {{Cytoarchitectonic Segmentation}} of {{Human Brain
  Areas}} with {{Self}}-supervised {{Siamese Networks}},''
\newblock in {\em {{MICCAI}}}, 2018, pp. 663--671.

\bibitem{Amunts2013}
K.~Amunts et~al.,
\newblock ``{{BigBrain}}: {{An Ultrahigh}}-{{Resolution 3D Human Brain
  Model}},''
\newblock {\em Science}, vol. 340, no. 6139, pp. 1472--1475, 2013.

\bibitem{Hadsell2006}
R.~Hadsell et~al.,
\newblock ``Dimensionality reduction by learning an invariant mapping,''
\newblock in {\em {{CVPR}}}, 2006, vol.~2, pp. 1735--1742.

\bibitem{Chen2020}
T.~Chen et~al.,
\newblock ``A {{Simple Framework}} for {{Contrastive Learning}} of {{Visual
  Representations}},''
\newblock in {\em {{ICML}}}, 2020.

\bibitem{Khosla2020}
P.~Khosla et~al.,
\newblock ``Supervised {{Contrastive Learning}},''
\newblock {\em arXiv:2004.11362 [cs, stat]}, 2020.

\bibitem{Ronneberger2015}
O.~Ronneberger et~al.,
\newblock ``U-net: {{Convolutional}} networks for biomedical image
  segmentation,''
\newblock in {\em {{MICCAI}}}, 2015, pp. 234--241.

\bibitem{Ioffe2015}
S.~Ioffe and C.~Szegedy,
\newblock ``Batch {{Normalization}}: {{Accelerating Deep Network Training}} by
  {{Reducing Internal Covariate Shift}},''
\newblock in {\em {{ICML}}}, 2015, pp. 448--456.

\bibitem{Pohlen2017}
T.~Pohlen et~al.,
\newblock ``Full-resolution residual networks for semantic segmentation in
  street scenes,''
\newblock in {\em {{arXiv}} Preprint}, 2017.

\bibitem{You2017}
Y.~You et~al.,
\newblock ``Large {{Batch Training}} of {{Convolutional Networks}},''
\newblock {\em arXiv:1708.03888 [cs]}, 2017.

\bibitem{Krause2018}
D.~Krause and P.~Th{\"o}rnig,
\newblock ``{{JURECA}}: {{Modular}} supercomputer at {{J\"ulich Supercomputing
  Centre}},''
\newblock {\em Journal of large-scale research facilities}, vol. 4, pp. 132,
  2018.

\bibitem{Paszke2019}
A.~Paszke et~al.,
\newblock ``{{PyTorch}}: {{An}} imperative style, high-performance deep
  learning library,''
\newblock in {\em {{NIPS}}}, 2019, pp. 8024--8035.

\bibitem{WardJr1963}
J.~H. Ward~Jr,
\newblock ``Hierarchical grouping to optimize an objective function,''
\newblock {\em Journal of the American statistical association}, vol. 58, no.
  301, pp. 236--244, 1963.

\bibitem{Hinton2003}
G.~E. Hinton and S.~T. Roweis,
\newblock ``Stochastic neighbor embedding,''
\newblock in {\em {{NIPS}}}, 2003, pp. 857--864.

\end{thebibliography}

\end{document}